\documentclass[a4paper,11pt]{article}
\usepackage{pos}

\title{Analyzing the $D^*D^*D^*$ system: Hexaquark states and the Efimov effect}
\author*[a]{Pablo G. Ortega}

\affiliation[a]{Departamento de Física Fundamental, Universidad de Salamanca, E-37008 Salamanca, Spain \\
Instituto Universitario de F\'isica
Fundamental y Matem\'aticas (IUFFyM), Universidad de Salamanca, E-37008 Salamanca, Spain}

\emailAdd{pgortega@usal.es}

\abstract{
In this work we investigate the possible emergence of Efimov states in the $D^*D^*D^*$ system with $J^P=0^-$ and isospin $I=\tfrac{1}{2}$, assuming the existence of the heavy partner of the $T_{cc}^+$, dubbed $T_{cc}^*$, near the $D^*D^*$ threshold as predicted by Heavy-Quark Spin Symmetry.
We find that $(I)J^P=(\frac{1}{2})0^-$ three-body bound states can be formed, with properties that suggest that the Efimov effect can be realised for reasonable values of the molecular probability and binding energy of the $T_{cc}^*$.}

\FullConference{10th International Conference on Quarks and Nuclear Physics (QNP2024)\\
8-12 July, 2024\\
Barcelona, Spain\\}

\begin{document}
\maketitle

\section{Introduction}

The \emph{Efimov effect} is a universal phenomenon first described by Vitaly Efimov in the 70's~\cite{Efimov:1970zz} which has been widely explored in nuclear, atomic and hadronic physics.
Briefly, when two particles are loosely bound as a consequence of short-range attractive forces, the system acquires \emph{universal} properties, which implies that the details of the interaction can be described by its $S$-wave scattering length $a_{\rm sc}$. Then, a long-range effective three-body interaction emerges, which is proportional to $1/\rho^2$, with $\rho$ the hyperspherical radius related to the distance between the three particles~\cite{Naidon:2016dpf,Braaten:2003he,Hammer:2010kp}. This effective three-body interaction generates an infinite number of three-body bound states with a discrete scale invariance. For three identical bosons of mass $m$ interacting through a short-range two-body potential, the effective potential is attractive. Then, as $a_{\rm sc}\to\pm\infty$, an infinite series of three-body bound states emerges with a scaling factor $\lambda=e^{\pi/|s_0|}\approx 22.7$. The binding energies of successive states follow the scaling law $\Delta E^{(n)} \to Q^2\Delta E^{(n+1)}$, where $Q\to\lambda$ in the unitarity limit.
In cases where the scattering length is large but finite, the spectrum is no longer infinite, though a few shallow Efimov states may appear if certain conditions are met~\cite{Naidon:2016dpf,Braaten:2004rn}.

In 2021, the LHCb Collaboration identified a new tetraquark-like state in the $D^0D^0\pi^+$ invariant mass spectrum, referred to as $T_{cc}^+$, with a quark content of $cc\bar u\bar d$~\cite{LHCb:2021vvq}. This resonance lies just below the $D^{+}D^0$ threshold, with a binding energy of $\delta m_{\rm pole}=(360\pm40^{+4}_{-0})$ keV/c$^2$~\cite{LHCb:2021auc}. Its scattering length is measured at $a{\rm sc}^{\rm LHCb}=-7.15(51)$ fm, and the Weinberg factor, which indicates the probability of finding a compact component in the wave function, is constrained as $Z<0.52(0.58)$ at 90(95)\% C.L. These characteristics are consistent with a state that has a substantial $DD^*$ molecular component (see Ref.~\cite{Chen:2022asf} for a review of the $T_{cc}^+$).
Heavy-Quark Spin Symmetry (HQSS) implies that the interaction of heavy mesons is insensitive to the heavy quark spin, so the $D^*D^*$ interaction mirrors the $DD^*$ interaction in the $(I)J^P=(0)1^+$ channel, aside from $1/M_Q$ corrections, which would point to a bound state in the $D^*D^*$ system as well, dubbed $T_{cc}^*$. The existence of the $T_{cc}^*$ has been predicted by several groups with binding energies on the order of a few MeV (see Ref.~\cite{Ortega:2024ecy} for references).

In this work, we investigate the universality of the $D^*D^*D^*$ meson system in the $J^P=0^-$ sector with isospin $I=\tfrac{1}{2}$, assuming that the isoscalar partner of the $T_{cc}^+$ exists just below the $D^*D^*$ threshold as predicted by HQSS~\cite{Neubert:1993mb}.
Should the $T_{cc}^*$ exist with a small binding energy, it would enhance the likelihood of Efimov states in this sector. If confirmed, this would mark the first observation of the Efimov effect in hadronic physics, making the study of this system particularly valuable. We focus on the $D^*D^*D^*$ system rather than the $DD^*D^*$ one, as working with identical bosons simplifies the calculations. For three identical bosons, the two-body wave function must be symmetric, and in the $J^P=0^-$ sector, all $D^*D^*$ pairs interact attractively, a condition necessary for the Efimov effect to occur.

The paper is organized as follows: After this introduction, Sec.~\ref{sec:model} briefly presents the theoretical framework. In section~\ref{sec:results} the results are analyzed and discussed.
For more details on the calculation, the reader is kindly referred to Ref.~\cite{Ortega:2024ecy}.

\section{Theoretical framework}\label{sec:model}

%\subsection{Two-body interaction}

The $D^*D^*$ two-body amplitude is obtained from the on-shell approximation of the Bethe-Salpeter equation~\cite{Nieves:1999bx}, $ {\cal T}_2^{-1}(s) = {\cal V}^{-1}(s) - {\cal G}(s)$,
where ${\cal V}(s)$ is the two-meson interaction and ${\cal G}$ is the relativistic two-meson loop function
regularized via a sharp cutoff~\cite{Oller:1998hw}. The value of the cutoff is taken between $\Lambda=0.5$ GeV and $1$ GeV. Using $m=\tfrac{1}{2}(m_{D^{*0}}+m_{D^{*\pm}})=2008.55$ MeV, the loop function can be written as

\begin{equation}\label{eq:loop2}
 {\cal G}(s)= \frac{1}{(4\pi)^2}\left\{ \sigma\log\frac{\sigma\sqrt{1+\tfrac{m^2}{\Lambda^2}}+1}{\sigma\sqrt{1+\tfrac{m^2}{\Lambda^2}}-1} -2\log\left[\frac{\Lambda}{m}\left(1+\sqrt{1+\tfrac{m^2}{\Lambda^2}}\right)\right] \right\}\,,
\end{equation}
with $\sigma=\sqrt{1-4m^2/s}$.

The $D^*D^*$ potential, ${\cal V}$, is a $I=0$ $S$-wave interaction where the $DD^*-D^*D^*$ coupled-channels effect and the finite width of the $D^*$ are neglected.
The effect of the possible $T_{cc}^*$ internal composition is evaluated by using a general energy-dependent contact potential~\cite{Montesinos:2023qbx}

\begin{equation}\label{eq:2bodypot}
 {\cal V}^{-1}(s) = C_0 -C_1\frac{1-{\cal P}}{{\cal P}} (s-m_*^2),
\end{equation}
where $C_0$ and $C_1$ are constants fixed to generate a $D^*D^*$ bound state with mass $m_*=2m-{\cal B}_2$, whose values are related to the loop function ${\cal G}(s)$ at $s=m_*^2$. The ${\cal P}$ parameter controls the molecular probability in the $T_{cc}^*$ state. Then, the 2-body binding energy ${\cal B}_2$ and the probability ${\cal P}$ are the only free parameters of the calculation.

%\subsection{Three-body amplitude}

For the three-body amplitude we will use the so-called \emph{ladder amplitude}~\cite{Hansen:2015zga,Jackura:2020bsk,Dawid:2023jrj}, a relativistic formalism where the three-to-three scattering is decomposed as

\begin{equation}
{\cal M}_3(\vec p_i,\vec p_f) = {\cal D}(\vec p_i,\vec p_f)+{\cal M}_{df,3}(\vec p_i,\vec p_f)
\end{equation}
Here, ${\cal D}$ is the ladder amplitude, which contains the sum over all possible pairwise interactions connected through a sequence of one-particle exchanges, and ${\cal M}_{df,3}$ includes all the contributions coming from short-range three-particle interactions.
In this study we fix ${\cal M}_{df,3}=0$, ignoring possible short-range three-body forces so the ladder amplitude emerges from two-body interactions alone.
The $\vec p_i$ ($\vec p_f$) is the initial (final) momentum of one of the particles, called the \emph{spectator}, which interacts with the remaining two particles, which are denoted as the \emph{dimer}.

In this work, all two-body subsystems interact in a relative $S$ wave. Additionally, the dimer-spectator system is also assumed to be in $L=0$, as this is expected to be the dominant partial wave~\cite{Bayar:2022bnc,Luo:2021ggs}. In fact, Ref.~\cite{Luo:2021ggs} examined the $D^*D^*D^*$ system in various configurations with $L\le 2$ and found that $S$-$D$ mixing had minimal impact compared to a purely $S$-wave calculation.

Hence, the  ladder amplitude ${\cal D}$ can be written as

\begin{equation}\label{eq2}
 {\cal D}(\vec p_i,\vec p_f) = -{\cal M}_2(p_i)G(\vec p_i,\vec p_f){\cal M}_2(p_f) -{\cal M}_2(p_i)\int \frac{d^3\vec q}{(2\pi)^32\omega(q)}G(\vec p_i,\vec q){\cal D}(\vec q,\vec p_f)\,,
\end{equation}
were $G$ is the long-range interaction between the dimer and the spectator and ${\cal M}_2$ is the two-body scattering amplitude. The dimer energy is fixed by the momentum of the spectator, $s_2(p)=(\sqrt{s}-\omega(p))^2-p^2$, with $\omega(p)=\sqrt{m^2+p^2}$ and $p=|\vec p|$. ${\cal M}_2$ can be related to the two-body $T$ matrix from the Bethe-Salpeter equation as ${\cal M}_2=-2\,{\cal T}_2$.
The ladder amplitude ${\cal D}$ is, then, projected to $S$ wave and numerically analyzed following the procedure of Ref.~\cite{Ortega:2024ecy,Dawid:2023jrj}.

\section{Results}\label{sec:results}

Our aim is to evaluate the existence of Efimov states in the $D^*D^*D^*$ system in the $(I)J^P=(\tfrac{1}{2})0^-$ sector assuming the existence of the HQSS partner of the $T_{cc}^+$, that's it, a bound $(I)J^P=(0)1^+$ $D^*D^*$ state, which has not been detected yet. The Efimov effect occurs when the two-body system is close to the resonant limit, so its appearance ultimately depends on the mass of the hypothetical $T_{cc}^*$. In order to explore the evolution of the Efimov states, we will take a selection of $T_{cc}^*$ binding energies from the literature, ${\cal B}_2=\{0.01,0.5,1.0,5.0\}$ MeV, and analyze the properties of the three-body states as a function of the $T_{cc}^*$ molecular content ${\cal P}$.

\begin{figure}[t!]
 \includegraphics[width=.45\textwidth]{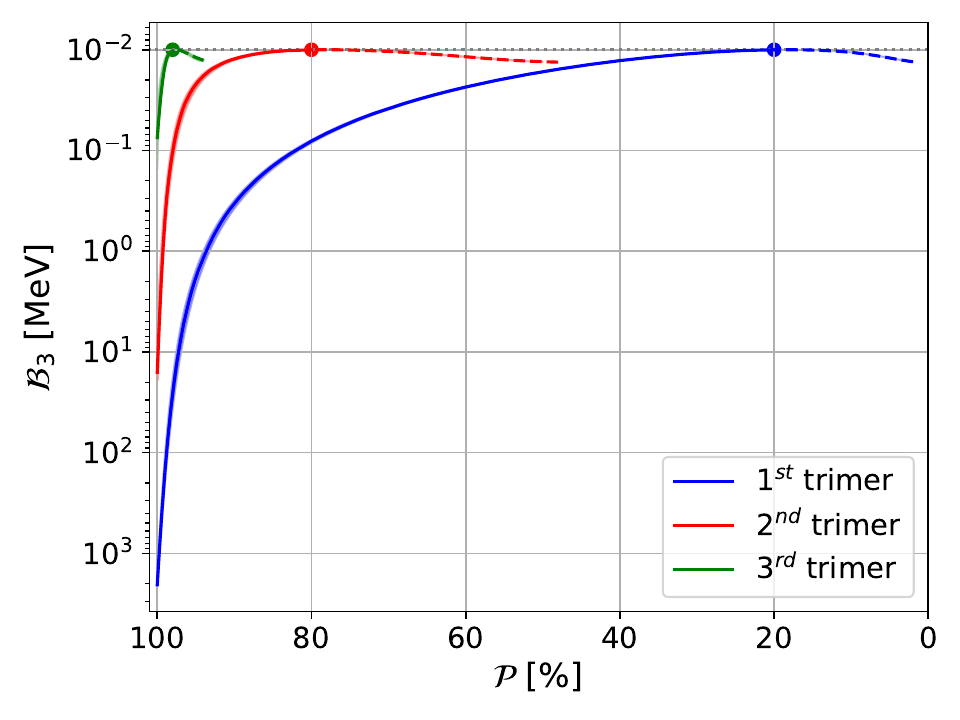}
 \includegraphics[width=.45\textwidth]{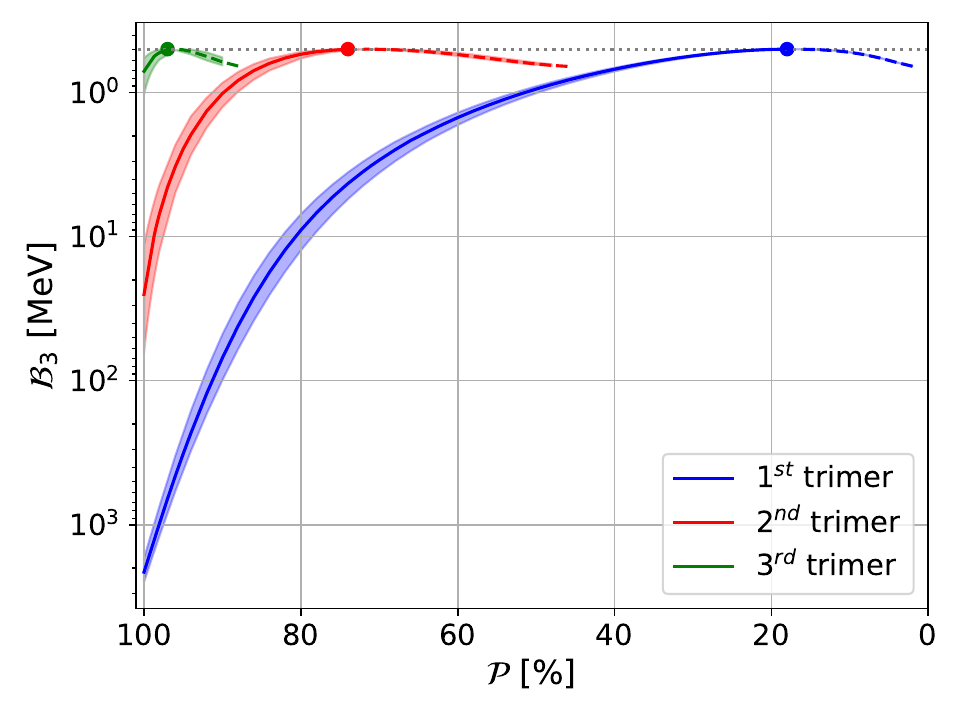}

 \includegraphics[width=.45\textwidth]{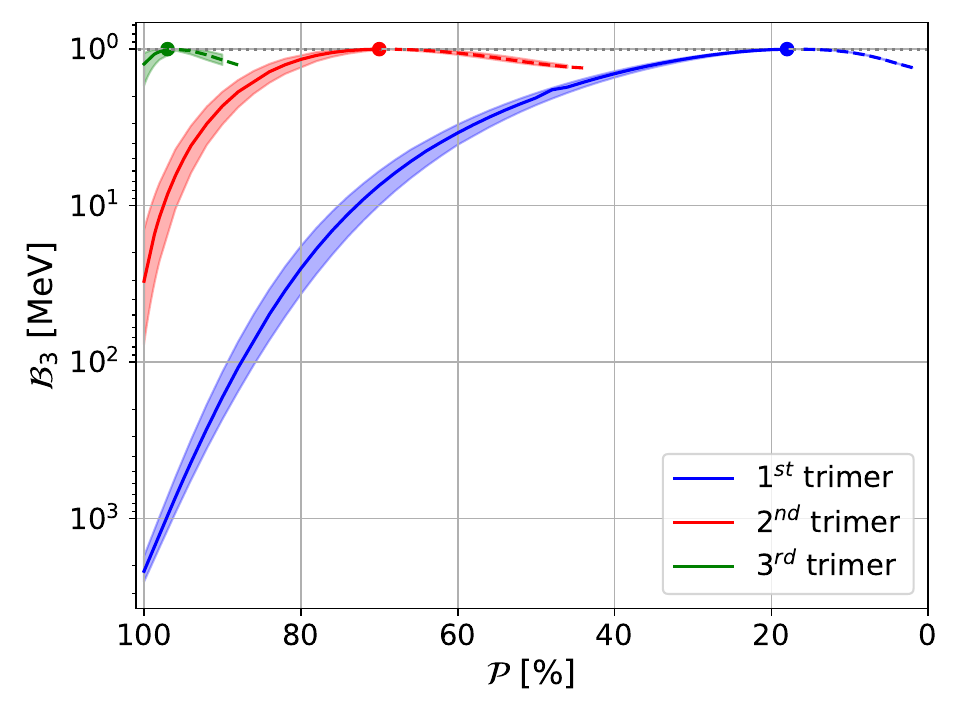}
 \includegraphics[width=.45\textwidth]{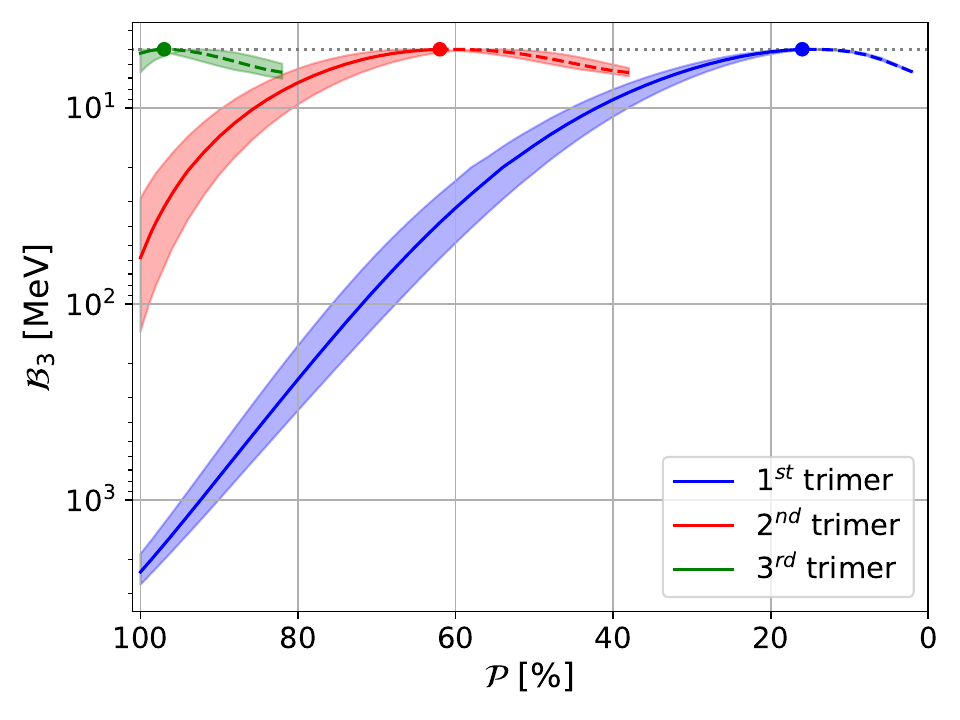}
 \caption{\label{fig:trimers} Binding energies of the first $D^*D^*D^*$ trimer masses (${\cal B}_3=E_3-3m$) for ${\cal B}_2=0.01$ MeV (upper left panel), $0.5$ MeV (upper right), $1$ MeV (lower left) and $5$ MeV (lower right panel) as a function of the $T_{cc}^*$ composition, ranging from a purely two-body molecular state (${\cal P}=100\%$) to a purely compact $T_{cc}^*$ state (${\cal P}=0\%$). The central lines show the results for $\Lambda=0.7$ GeV cutoff in the two-body amplitude. The color error bands indicate the results for the cutoff range $\Lambda=[0.5,1]$ GeV. Solid lines represent bound states, whereas dashed lines represent virtual states. The dot marks the value of ${\cal P}$ where the pole changes the Riemann sheet. The dotted horizontal gray line shows the two-body binding energy ${\cal B}_2$, which acts as the threshold for the trimer states. Figures taken from Ref.~\cite{Ortega:2024ecy}. }
\end{figure}

Using the two-body potential of Eq.~\eqref{eq:2bodypot}, up to three trimers are found~\cite{Ortega:2024ecy}. Their binding energies, however, evolve with the molecular probability of the $T_{cc}^*$, as the coupling with the $D^*D^*$ system changes with ${\cal P}$. In Fig.~\ref{fig:trimers} we show the $D^*D^*D^*$ trimers for  the four binding energies under evaluation, ${\cal B}_2=0.01$ MeV, $0.5$ MeV, $1$ MeV and $5$ MeV. The trimers acquire a deeper binding energy the larger ${\cal P}$ and ${\cal B}_2$. For ${\cal P}>20\%$ the first trimer emerges as a bound state, while the second emerges between $60\%$ and $80\%$, depending on ${\cal B}_2$. The third trimer needs an almost pure $D^*D^*$ molecular component in the $T_{cc}^*$, ${\cal P}>96\%$, but its detection will be challenging due to its small binding energy.

\begin{figure}[t!]
\begin{center}
 \includegraphics[width=.48\textwidth]{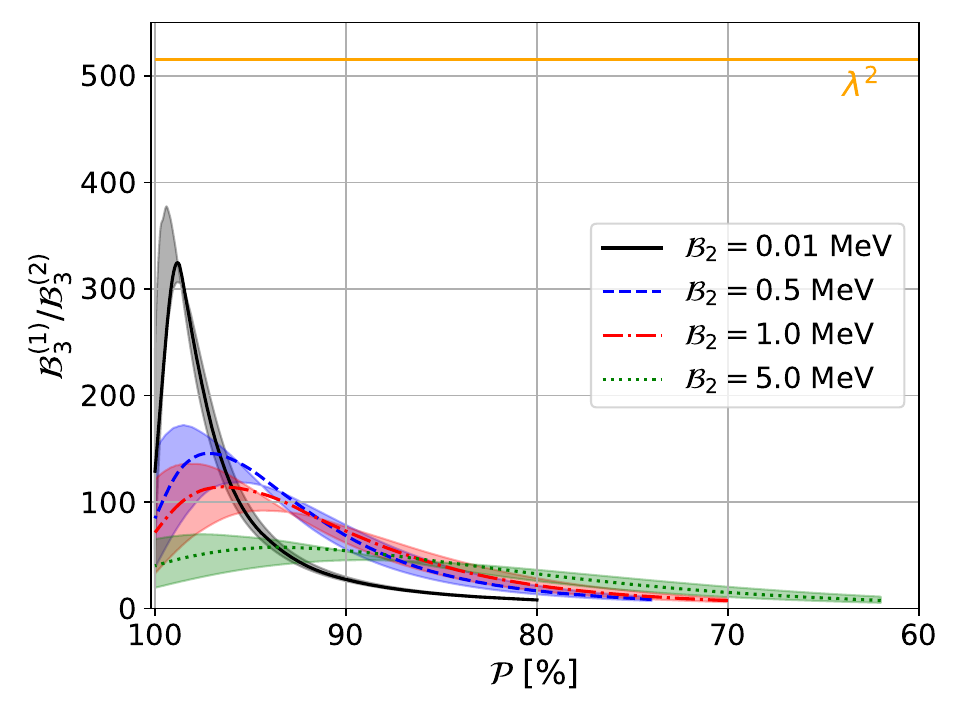}
\end{center}
 \caption{\label{fig:ratios} Ratio of the first to second trimer binding energies for different ${\cal B}_2$ as a function of the $T_{cc}^*$ composition. The orange horizontal line represents the Efimov scaling factor  $\lambda^{2}\sim 515$ at the unitary limit $a_{\rm sc}\to\infty$. A cutoff of $\Lambda=0.7$ GeV has been used in Eq.~\eqref{eq:loop2}. The color error bands indicate the results for the cutoff range $\Lambda=[0.5,1]$ GeV.}
\end{figure}

Figure~\ref{fig:ratios} shows the ratios of the binding energies for the first and second trimer. The results show that the ratios draw a peak structure at small values of ${\cal P}$, the height of which decreases as ${\cal B}_2$ increases. The peak approaches the Efimov scaling law, $\lambda=e^{2\pi/|s_0|}\sim 515$, as ${\cal B}_2$ decreases and reaches this value when ${\cal B}_2 \to 0$. This deviation from the universal $\lambda^2$ value arises because the system is not exactly at the unitarity limit~\cite{Dawid:2023kxu}. In addition, possible compact components in the $T_{cc}^*$ wave function, which mix with the $D^*D^*$ pairs, further alter this scaling.

As a summary, this work has analyzed the $D^*D^*D^*$ system in the $(I)J^P=(\tfrac{1}{2})0^-$ sector and found evidence that the Efimov effect can indeed form.
Needless to say, the discovery of the Efimov effect in systems involving charmed mesons would be groundbreaking and would greatly advance our understanding of multimeson states, making this system a promising candidate for further investigation.

%%%%%%%%%%%%%%%%%%%%%%%%%%%%%%%%%%%%%%%%%%%%%%%%%%%%%%%%%%%%%%%%%%%%%%%%%%%%%%%%
%%%%%%%%%%%%%%%%%%%%%%%%%%%%%%%%%%%%%%%%%%%%%%%%%%%%%%%%%%%%%%%%%%%%%%%%%%%%%%%%

\section{Acknowledgments}
This work has been partially funded by
EU Horizon 2020 research and innovation program, STRONG-2020 project, under grant agreement no. 824093;
Junta de Castilla y León EDU/841/2024 program under grant agreement no. SA091P24; and
Ministerio Espa\~nol de Ciencia e Innovaci\'on under grant Nos. PID2019-105439GB-C22 and PID2022-141910NB-I00.

\bibliographystyle{apsrev4-1} % Tell bibtex which bibliography style to use
\bibliography{Biblio}

\end{document}